\documentclass[doublespacing]{elsart}

\newcommand {\pt}	{p_{T}}
\newcommand {\phit}	{\phi_{t}}
\newcommand {\phis}	{\phi_{s}}
\newcommand {\psiEP}	{\psi_{EP}}
\newcommand {\psiRP}	{\psi_{RP}}
\newcommand {\dpsi}	{\Delta\psi}
\newcommand {\dphi}	{\Delta\phi}
\newcommand {\Nt}	{N_{t}}
\newcommand {\NtR}	{\Nt^{(R)}}
\newcommand {\vt}[1]	{v_{#1}^{(t)}}
\newcommand {\mean}[1]	{\langle#1\rangle}
\newcommand {\be}	{\begin{equation}}
\newcommand {\ee}	{\end{equation}}
\newcommand {\bea}	{\begin{eqnarray}}
\newcommand {\eea}	{\end{eqnarray}}

\usepackage{psfig}
\usepackage{lineno}
\linenumbers

\begin{document}
\begin{frontmatter}


\title{Anisotropic flow background to jet-correlation relative to reaction plane: a revisit of the mathematical framework}


\author{Joshua Konzer}
\ead{jrkonzer@purdue.edu}
\author{Fuqiang Wang}
\ead{fqwang@purdue.edu}
\address{Department of Physics, Purdue University, 525 Northwestern Ave., West Lafayette, IN 47907, USA}

\begin{abstract}
We derive the flow-background formula for jet-correlation analysis with high $\pt$ trigger particles in {\em any} azimuth window relative to reaction plane, extending the mathematic framework of previous study in~\cite{Jana}.
\end{abstract}

\end{frontmatter}

\maketitle

Strong interaction of hard-scattering partons with the high density medium created in relativistic heavy-ion collisions results in modification to jet-correlation with high transverse momentum ($\pt$) trigger particles due to jet energy loss~\cite{xnwang,wp1,wp2}. The away-side correlation opposite to the trigger particle was found to be strongly modified~\cite{prl95,phenix}; the correlation structure is significantly broadened, and the lost energy is dissipated into low $\pt$ particles~\cite{prl95} and appears to excite Mach cone shock sound waves in the created medium~\cite{Machcone}. The near-side correlation was found to be extended to large speudo-rapidity~\cite{prl95,ridge,Levente,phobos}. The long-ranged correlation is dubbed as the ``ridge". The physics of the ridge is not well understood.

In turn, jet-correlation measurement with high $\pt$ trigger particles has become a powerful tool to study medium properties~\cite{wp1,wp2}. Jet-correlation relative to reaction plane provides further capability by exploiting the non-sphericity of the initial nuclear overlap region~\cite{Aoqi}. It was found that the ridge magnitude drops significantly with trigger particles from in-plane to out-of-plane in medium central Au+Au collisions~\cite{Aoqi}. In particular, with the trigger particle pointing at an angle relative to the reaction plane in a non-central collision, the medium thicknesses along the directions at the two sides of the away-side parton are different. This asymmetry in the medium thickness should have diffeerent effects on the gluons radiated off the away-side parton and can result in asymmetric correlation functions. A recent study~\cite{Jia} explored such possibilities.

Recently Chiu and Hwa~\cite{Hwa} proposed a physics mechanism for the observed ridge as being a result of interactions between radiated gluons and medium flow. They argue that when jet propogation is aligned with the medium flow direction, small-angle radiated gluons are collimated with the jet direction, while the radiated gluons are pushed away from the jet direction when jet propagation is perpendicular to the medium flow direction~\cite{Hwa}. This model of gluon emission correlated with medium flow can explain the observed descrease of ridge magnitude from in-plane to out-of-plane~\cite{Aoqi}. Furthermore, the model predicts asymmetric azimuthal correlation functions for trigger particles in quadrants I+III and quadrants II and IV, separately~\cite{Hwa}. Such asymmetry may not be restricted only to this particular model implementation, but a more general feature from interactions between jets and medium flow (e.g., see Refs.~\cite{Voloshin,Shuryak,McLerran1,McLerran2,Longacre,Takahashi}). Clearly, experimental test of this physics mechanism is important to further our understanding of jet-medium interaction. 

One important aspect of jet-correlation analysis is the subtraction of combinatorial background which itself has an azimuthal dependence due to anisotropic flow of background particles. The formulism of this flow background is given in Ref.~\cite{Jana} for trigger particles within a restricted azimuthal range relative to reaction plane, summed over all four quadrants. Due to reflection symmetry, flow backgrounds in quadrants I and III are identical and so are those in quadrants II and IV, but flow backgrounds in quadrants I+III and II+IV are not the same. The flow-background formula given in Ref.~\cite{Jana}, summed over all four quadrants, cannot be readily used for jet-correlation analysis separating trigger particles in quadrants I+III and II+IV. In this note we derive the needed formula.

Due to event anisotropy, particle azimuthal distribution is given by
\be
\frac{dN}{d\phi}=\frac{N}{2\pi}\left[1+\sum_{k=1}^{\infty}2v_k\cos k(\phi-\psiRP)\right]
\ee
where $\psiRP$ is the reaction plane azimuthal angle, and $v_k$ is the $k$th harmonic coefficient. Trigger-associated particle pair distribution is given by
\bea
\frac{d^4N}{d\psiRP d\phit d\phi d\dphi}&=&\frac{1}{2\pi}\frac{\Nt}{2\pi}\left[1+\sum_{k=1}^{\infty}2\vt{k}\cos k(\phit-\psiRP)\right]\times\nonumber\\
&&\frac{N}{2\pi}\left[1+\sum_{k=1}^{\infty}2v_k\cos k(\phi-\psiRP)\right]\delta(\dphi-\phi+\phit)
\eea
Integrating over trigger particle azimuthal angle $\phit$ within a slice $|\phit-(\psiRP+\phis)|<c$, and associated particle $\phi$ and $\psiRP$ over full $2\pi$, we have
\bea
\frac{dN}{d\dphi}=\frac{\Nt N}{(2\pi)^3}\int_{0}^{2\pi}d\psiRP\int_{\psiRP+\phis-c}^{\psiRP+\phis+c}&&d\phit\left[1+\sum_{k=1}^{\infty}2\vt{k}\cos k(\phit-\psiRP)\right]\nonumber\\
&&\times\left[1+\sum_{k=1}^{\infty}2v_k\cos k(\dphi+\phit-\psiRP)\right]
\eea

However, we do not know the real reaction plane $\psiRP$, but only the measured event plane $\psiEP$, which is smeared from $\psiRP$ by probability function $\rho(\dpsi)$ where $\dpsi=\psiEP-\psiRP$ and $\int_{0}^{2\pi}\rho(\dpsi)d\dpsi\equiv 1$. Experimentally trigger particles are selected within slice $|\phit-(\psiEP+\phis)|<c$, hence
\bea
\frac{dN}{d\dphi}&=&\frac{\Nt N}{(2\pi)^3}\int_{0}^{2\pi}d\psiRP\int_{0}^{2\pi}\rho(\dpsi)d\dpsi\int_{\psiRP+\dpsi+\phis-c}^{\psiRP+\dpsi+\phis+c}d\phit\times\nonumber\\
&&\left[1+\sum_{k=1}^{\infty}2\vt{k}\cos k(\phit-\psiRP)\right]\times\left[1+\sum_{k=1}^{\infty}2v_k\cos k(\dphi+\phit-\psiRP)\right]
\eea
The integrand is
\bea
1&+&\sum_{k=1}^{\infty}2\vt{k}\cos k(\phit-\psiRP)+\sum_{k=1}^{\infty}2v_k\cos k(\dphi+\phit-\psiRP)+\nonumber\\
&&\sum_{j=1}^{\infty}\sum_{i=1}^{\infty}2\vt{i}v_j\left\{\cos\left[(j+i)(\phit-\psiRP)+j\dphi\right]+\cos\left[(j-i)(\phit-\psiRP)+j\dphi\right]\right\}
\eea
Integrating over $\phit$, we obtain
\bea
\frac{dN}{d\dphi}&=&
\frac{2c\Nt N}{(2\pi)^2}\int_{0}^{2\pi}\rho(\dpsi)d\dpsi
\left\{1+
\sum_{k=1}^{\infty}2\vt{k}\frac{\sin kc}{kc}\cos k(\phis+\dpsi)+\right.\nonumber\\&&
\sum_{k=1}^{\infty}2v_k\frac{\sin kc}{kc}\cos k(\dphi+\phis+\dpsi)+\sum_{k=1}^{\infty}2\vt{k}v_k\cos k\dphi+\nonumber\\&&
\sum_{i=1}^{\infty}\sum_{j=1}^{\infty}2\vt{i}v_j\frac{\sin(j+i)c}{(j+i)c}\cos\left[(j+i)(\phis+\dpsi)+j\dphi\right]+\nonumber\\&&
\left.\sum_{j=1}^{\infty}\sum_{i=1,i\neq j}^{\infty}2\vt{i}v_j\frac{\sin(j-i)c}{(j-i)c}\cos\left[(j-i)(\phis+\dpsi)+j\dphi\right]
\right\}
\eea
Taking $\int_{0}^{2\pi}\rho(\dpsi)d\dpsi\cos n\dphi\equiv\mean{\cos n\dphi}$ and $\int_{0}^{2\pi}\rho(\dpsi)d\dpsi\sin n\dphi=0$, we have
\bea
\frac{dN}{d\dphi}&=&
\frac{2c\Nt N}{(2\pi)^2}\left\{1+
\sum_{k=1}^{\infty}2\vt{k}\cos k\phis\frac{\sin kc}{kc}\mean{\cos k\dpsi}+\right.\nonumber\\&&
\sum_{k=1}^{\infty}2v_k\cos k(\dphi+\phis)\frac{\sin kc}{kc}\mean{\cos k\dpsi}+\sum_{k=1}^{\infty}2\vt{k}v_k\cos k\dphi+\nonumber\\&&
\sum_{j=1}^{\infty}\sum_{i=1}^{\infty}2\vt{i}v_j\cos\left[(j+i)\phis+j\dphi\right]\frac{\sin(j+i)c}{(j+i)c}\mean{\cos(j+i)\dpsi}+\nonumber\\&&
\left.\sum_{j=1}^{\infty}\sum_{i=1,i\neq j}^{\infty}2\vt{i}v_j\cos\left[(j-i)\phis+j\dphi\right]\frac{\sin(j-i)c}{(j-i)c}\mean{\cos(j-i)\dpsi}
\right\}
\eea
The two $\sum\sum$ terms can be rewritten into:
\bea
&&\sum_{j=1}^{\infty}2v_j\sum_{k=j+1}^{\infty}\vt{k-j}\cos(k\phis+j\dphi)\frac{\sin kc}{kc}\mean{\cos k\dpsi}+\nonumber\\
&&\sum_{j=1}^{\infty}2v_j\sum_{k=1-j,k\neq0}^{\infty}\vt{k+j}\cos(k\phis-j\dphi)\frac{\sin kc}{kc}\mean{\cos k\dpsi}\nonumber\\
&=&\sum_{j=1}^{\infty}2v_j\sum_{k=j+1}^{\infty}\vt{k-j}\cos(k\phis+j\dphi)\frac{\sin kc}{kc}\mean{\cos k\dpsi}+\nonumber\\
&&\sum_{j=1}^{\infty}2v_j\left(\sum_{k=1}^{\infty}\vt{k+j}+\sum_{k=1-j}^{-1}\vt{k+j}\right)\cos(k\phis-j\dphi)\frac{\sin kc}{kc}\mean{\cos k\dpsi}\nonumber\\
&=&\sum_{j=1}^{\infty}2v_j\left(\sum_{k=j+1}^{\infty}\vt{k-j}+\sum_{k=1}^{j-1}\vt{j-k}\right)\cos(k\phis+j\dphi)\frac{\sin kc}{kc}\mean{\cos k\dpsi}+\nonumber\\
&&\sum_{j=1}^{\infty}2v_j\sum_{k=1}^{\infty}\vt{k+j}\cos(k\phis-j\dphi)\frac{\sin kc}{kc}\mean{\cos k\dpsi}\nonumber\\
&=&\sum_{j=1}^{\infty}2v_j\sum_{k=1,k\neq j}^{\infty}\vt{|k-j|}\cos(k\phis+j\dphi)\frac{\sin kc}{kc}\mean{\cos k\dpsi}+\nonumber\\
&&\sum_{j=1}^{\infty}2v_j\sum_{k=1}^{\infty}\vt{k+j}\cos(k\phis-j\dphi)\frac{\sin kc}{kc}\mean{\cos k\dpsi}
\eea
Thus
\bea
\frac{dN}{d\dphi}&=&
\frac{2c\Nt N}{(2\pi)^2}\left\{1+
\sum_{k=1}^{\infty}2\vt{k}\cos k\phis\frac{\sin kc}{kc}\mean{\cos k\dpsi}+\right.\nonumber\\&&
\sum_{k=1}^{\infty}2v_k\cos k(\dphi+\phis)\frac{\sin kc}{kc}\mean{\cos k\dpsi}+
\sum_{k=1}^{\infty}2\vt{k}v_k\cos k\dphi+\nonumber\\&&
\sum_{j=1}^{\infty}2v_j\sum_{k=1,k\neq j}^{\infty}\vt{|k-j|}\cos(k\phis+j\dphi)\frac{\sin kc}{kc}\mean{\cos k\dpsi}+\nonumber\\&&
\left.\sum_{j=1}^{\infty}2v_j\sum_{k=1}^{\infty}\vt{k+j}\cos(k\phis-j\dphi)\frac{\sin kc}{kc}\mean{\cos k\dpsi}\right\}
\eea
The 3rd and 5th term can be combined if using convention $\vt{0}=1$, yielding
\bea
\frac{dN}{d\dphi}&=&
\frac{2c\Nt N}{(2\pi)^2}\left\{1+
\sum_{k=1}^{\infty}2\vt{k}\cos k\phis\frac{\sin kc}{kc}\mean{\cos k\dpsi}+
\sum_{k=1}^{\infty}2\vt{k}v_k\cos k\dphi+\right.\nonumber\\&&
\left.\sum_{j=1}^{\infty}2v_j\sum_{k=1}^{\infty}\frac{\sin kc}{kc}\mean{\cos k\dpsi}\left[\vt{|k-j|}\cos(k\phis+j\dphi)+\vt{k+j}\cos(k\phis-j\dphi)\right]
\right\}
\eea

The number of trigger particles within the given slice is 
\bea
\NtR&=&\frac{\Nt}{(2\pi)^2}\int_{0}^{2\pi}d\psi\int_{0}^{2\pi}\rho(\dpsi)d\dpsi\int_{\psiRP+\dpsi+\phis-c}^{\psiRP+\dpsi+\phis+c}d\phit\left[1+\sum_{k=1}^{\infty}2\vt{k}\cos k(\phit-\psiRP)\right]\nonumber\\
&=&\frac{2c\Nt}{(2\pi)^2}\int_{0}^{2\pi}d\psi\int_{0}^{2\pi}\rho(\dpsi)d\dpsi\left[1+\sum_{k=1}^{\infty}2\vt{k}\frac{\sin kc}{kc}\cos k(\dpsi+\phis)\right]\nonumber\\
&=&\frac{2c\Nt}{2\pi}\left(1+\sum_{k=1}^{\infty}2\vt{k}\cos k\phis\frac{\sin kc}{kc}\mean{\cos k\dpsi}\right)
\eea
Normalized by the number of trigger particles, the correlation function becomes
\bea
\frac{1}{\NtR}\frac{dN}{d\dphi}&=&\frac{N}{2\pi}\left(1+\sum_{i=1}^{\infty}2v_i\times\right.\nonumber\\&&
\left.\frac{\vt{i}\cos i\dphi+\displaystyle{\sum_{k=1}^{\infty}}\frac{\sin kc}{kc}\mean{\cos k\dpsi}\left[\vt{|k-i|}\cos(i\dphi+k\phis)+\vt{k+i}\cos(i\dphi-k\phis)\right]}
{1+\displaystyle{\sum_{k=1}^{\infty}}2\vt{k}\cos k\phis\frac{\sin kc}{kc}\mean{\cos k\dpsi}}
\right)
\eea
Using shorthand notation
\be
R_k=\frac{\sin kc}{kc}\mean{\cos k\dpsi},
\ee
we rewrite our result into
\bea
\frac{1}{\NtR}\frac{dN}{d\dphi}&=&\frac{N}{2\pi}\left(1+\sum_{i=1}^{\infty}2v_i\times\right.\nonumber\\&&
\left.\frac{\vt{i}\cos i\dphi+\displaystyle{\sum_{k=1}^{\infty}}R_k\left[\vt{|k-i|}\cos(i\dphi+k\phis)+\vt{k+i}\cos(i\dphi-k\phis)\right]}
{1+\displaystyle{\sum_{k=1}^{\infty}}2\vt{k}R_k\cos k\phis}
\right)
\label{eq:result}
\eea

At mid-rapidity, odd harmonics vanish. Thus the result is invariant under $\phis\rightarrow\phis+\pi$. The result is different under $\phis\rightarrow-\phis$; both $\frac{dN(\phis)}{d\dphi}$ and $\frac{dN(-\phis)}{d\dphi}$ are asymmetric about $\dphi=0$, but the average of the two,
\be
\frac{1}{\NtR}\frac{dN}{d\dphi}=\frac{N}{2\pi}\left(1+\sum_{i=\mathrm{even}}^{\infty}2v_i
\frac{\vt{i}+\displaystyle{\sum_{k=\mathrm{even}}^{\infty}}\left(\vt{|k-i|}+\vt{k+i}\right)R_k\cos k\phis}
{1+\displaystyle{\sum_{k=\mathrm{even}}^{\infty}}2\vt{k}R_k\cos k\phis}\cos i\dphi
\right)\label{eq:symm}
\ee
is symmetric. Eq.~\ref{eq:symm} recovers the result in Ref.~\cite{Jana}.

Keeping terms up to $v_2v_4$ and $v_6$, Eq.~\ref{eq:result} becomes
\bea
\frac{1}{\NtR}\frac{dN}{d\dphi}&=&
\frac{N}{2\pi}\left\{1+
\frac{1}{1+\displaystyle{\sum_{k=2,4,6}}2\vt{k}R_k\cos k\phis}\times\right.\nonumber\\&&
\left.\left(
\begin{array}{c}
2v_2R_2\cos2(\dphi+\phis)+2v_4R_4\cos4(\dphi+\phis)+\\
2v_6R_6\cos6(\dphi+\phis)+2v_2\vt{2}\left[\cos2\dphi+R_4\cos2(\dphi+2\phis)\right]\\
+2v_2\vt{4}\left[R_2\cos2(\dphi-\phis)+R_6\cos2(\dphi+2\phis)\right]+\\
2\vt{2}v_4\left[R_2\cos2(2\dphi+\phis)+R_6\cos2(2\dphi+3\phis)\right]
\end{array}
\right)\right\}
\label{eq:approx}
\eea

\begin{figure*}[hbt]
\centerline{
\psfig{file=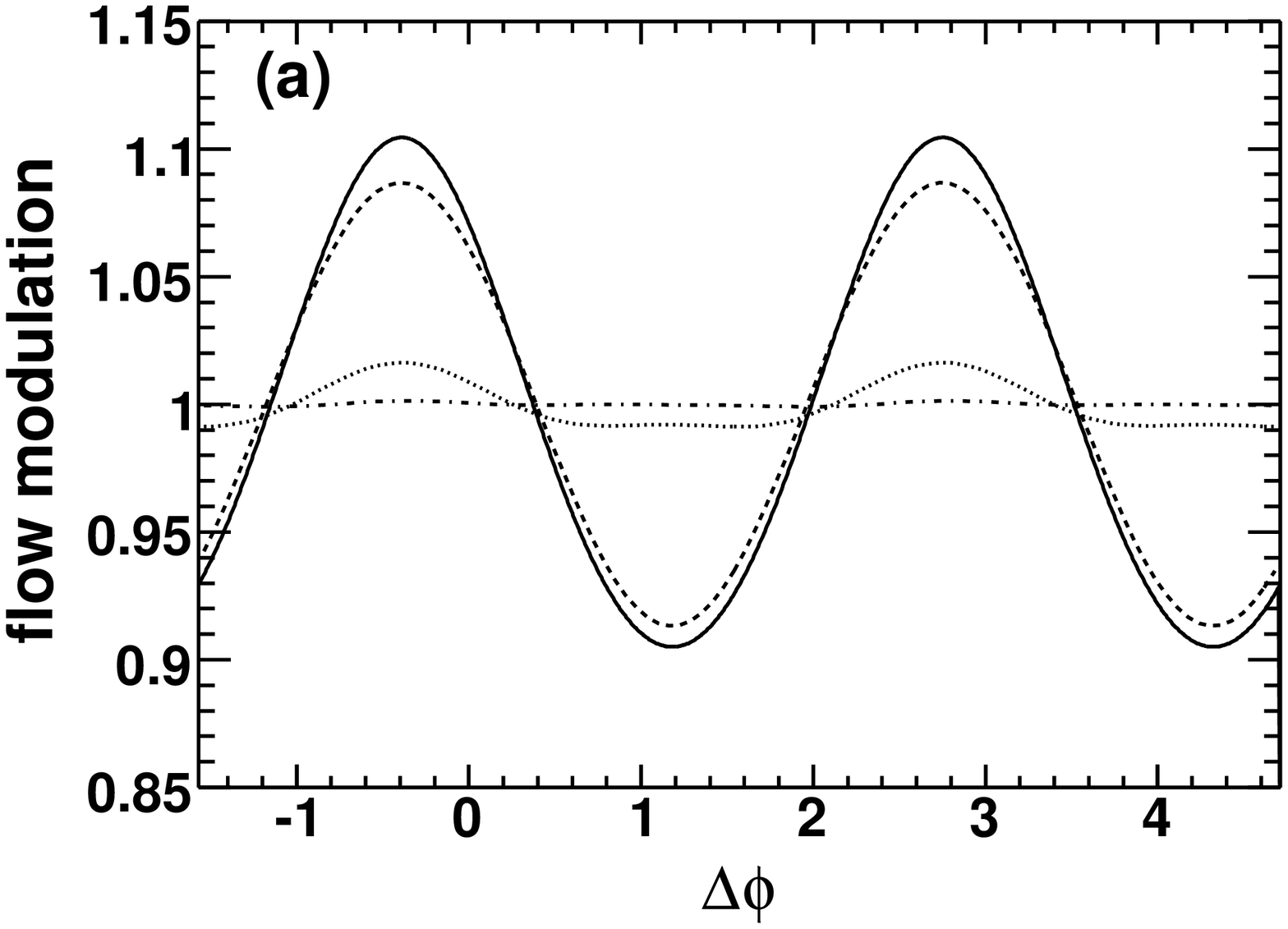,width=0.5\textwidth}
\psfig{file=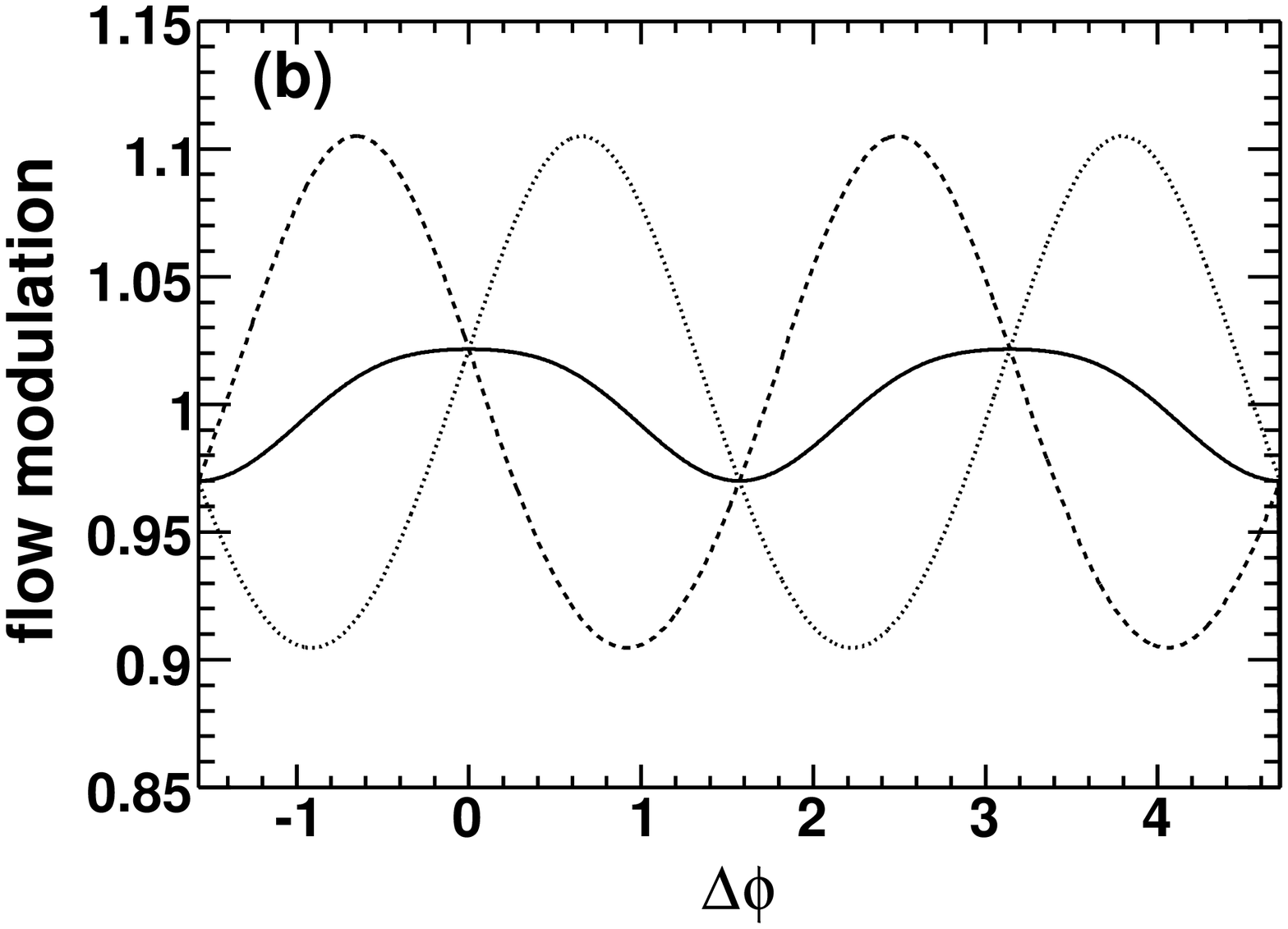,width=0.5\textwidth}
}
\caption{(a) Flow modulation, the quantity in the curly brackets in Eq.~(\ref{eq:approx} for trigger particle azimuthal angle range $15^{\circ}<\phit<30^{\circ}$ (solid curve). Typical elliptic flow magnitudes of $\vt{2}=0.10$ and $v_2=0.05$, and $\vt{4}=(\vt{2})^2, \vt{6}=(\vt{2})^3, v_4=v_2^2$ and $v_6=v_2^3$ are used and event plane resolutions are taken to be $\mean{\cos2\dpsi}=\mean{\cos4\dpsi}=\mean{\cos6\dpsi}=1$. The dashed, dotted, and dash-dotted curves are the individual contributions of the order $v_2$, $v_4$, and $v_6$, respectively. (b) Flow modulations for ranges $30^{\circ}<\phit<45^{\circ}$ (dashed curve), $-45^{\circ}<\phit<-30^{\circ}$ (dotted curve), and the combined range $30^{\circ}<|\phit|<45^{\circ}$ (solid curve).}
\label{fig}
\end{figure*}

For illustration we calcualte the flow modulation, the quantity in the curly brackets in Eq.~(\ref{eq:approx}), for typical elliptic flow magnitudes of $\vt{2}=0.10$ and $v_2=0.05$, and $\vt{4}=(\vt{2})^2, \vt{6}=(\vt{2})^3, v_4=v_2^2$ and $v_6=v_2^3$. We shall take the event plane resolutions to be $\mean{\cos2\dpsi}=1, \mean{\cos4\dpsi}=1$, and $\mean{\cos6\dpsi}=1$ (i.e., perfect event plane reconstruction). For smaller resolutions, the flow modulation is reduced. The calculated flow modulation is shown in Fig.~\ref{fig}(a) in the solid curve for trigger particle azimuthal angle range $15^{\circ}<\phit<30^{\circ}$ (i.e., $\phis=22.5^{\circ}, c=7.5^{\circ}$). Also shown are the individual contributions of the order $v_2$, $v_4$, and $v_6$, in dashed, dotted, and dash-dotted curves, respectively. In Fig.~\ref{fig}(b) we show the flow modulations for ranges $30^{\circ}<\phit<45^{\circ}$ (dashed curve), $-45^{\circ}<\phit<-30^{\circ}$ (dotted curve), and the combined range $30^{\circ}<|\phit|<45^{\circ}$ (solid curve). 

In summary, we have derived the flow-background formula for jet-correlation analysis with high $\pt$ trigger particles in {\em any} azimuth window relative to reaction plane, extending the mathematic framework of previous study in~\cite{Jana}. Our main result is in Eq.~(\ref{eq:result}). An approximation up to the order of $v_2v_4$ (and $v_6$) is given in Eq.~(\ref{eq:approx}).

{\em Acknowledgment--} This work is supported by U.S. Department of Energy under Grant DE-FG02-88ER40412.


\end{document}